\documentclass[reqno]{amsart}

\usepackage[english]{babel}
\usepackage{graphicx}
\usepackage{textcomp}
\usepackage{amsmath}
\usepackage{amssymb}
\usepackage{latexsym}
\newtheorem{definition}{Definition}

\newtheorem{theorem}{Theorem}
\begin{document}
%======================================================================
%\cleanbegin
%======================================================================
\title[Arbitrage hedging strategy and the volatility smile]
{Arbitrage hedging strategy  and one more  explanation of the
volatility smile}

%on the markets with assets depending on the same random factor

%\author{\textbf{S.Albeverio, A.Korshunova, O.Rozanova}}% List of authors with the
%{Mathematics and Mechanics Department, Lomonosov  Moscow State University, Russia}                 % same affiliation,
                                                  % lecturer given in bold
%{rozanova@mech.math.msu.su}                                   % e-mail of lecturer
%
%                                                 % it is important not to have
%                                                 % blank lines between the
%                                                 % \author and \coauthor
%
                                     % use such line
\author[Martynov, Rozanova]{Mikhail Martynov $^{1}$, Olga Rozanova $^{2}$}

%\author[label1,label2]{}
\address[$^{1}$, $^{2}$]{Mathematics and Mechanics Faculty, Moscow State University, Moscow
119992, Russia}

\thanks {Supported by the special program of the
Ministry of Education of the Russian Federation "The development of
scientific potential of the Higher School", project 2.1.1/1399.}

\email[$^{1}$]{mikhailmartynov@gmail.com}
\email[$^{2}$]{rozanova@mech.math.msu.su}

%\subjclass {35L65; 35L67}

                                                  % if e-mail none available
\maketitle

\begin{abstract}
We present an explicit hedging strategy, which enables to prove
arbitrageness  of market incorporating  at least two assets
depending on the same random factor.  The implied Black-Scholes
volatility, computed taking into account the  form of the graph of
the option price, related to our strategy, demonstrates the
"skewness" inherent to the observational data.
\end{abstract}

%------------------------------------------------------------------

\section{Introduction}

This work is an extension of the paper \cite{Martynov}, where the
arbitrage strategy was constructed for a market with pure correlated
assets.

  Assume that on a market
there exist at least two assets with prices
 $S_1(t)$ and $S_2(t)$ which are random processes dependent on the same Brownian motion.
 It is known that the market with the asset cannot be arbitrage-free.
  In \cite{Bjork} the principle is formulated that the market is arbitrage-free if and only
  if the number of  traded assets (excluding the riskless one) does not exceed the number of sources of
  randomness.
  In the same book it is shown by means of the martingale approach  that the market
  including several risky assets with prices given by the $\rm It\hat{o}$ processes is arbitrage-free if and only if
  the number of independent Wiener processes is equal or greater then  the number of the risky assets.

  This theoretical result correlates with well known condition for
  arbitrage:
  two assets with identical cash flows do not trade at the same
  price (e.g.\cite{Arbitrage}).

In the present paper we prove  arbitrageness of the market with
assets dependent on the same random factors in an alternative way
and construct an explicit hedging strategy using a mathematical
tool, describing a formation of contrast structures of step type in
solutions of semi-linear parabolic equations.

For the sake of definiteness we consider the call option.
\section{Initial-boundary problem for an analog of the Black- Scholes equation}

Let us consider a financial instrument with the price  $V = V(S_1,
S_2, t)$ dependent on prices  $S_1, S_2$ of two different assets and
assume that the market is arbitrage-free.

Assume that the prices of the assets are given as follows:
$$
dS_{1} = \mu_{1} S_{1} dt + \sigma_{1} S_{1} dW, \quad dS_{2} =
\mu_{2} S_{2} dt + \sigma_{2} S_{2} dW.
$$
We mean that the assets are different if their prices satisfy
geometrical Brownnian motions
 having at least different parameters of volatility
( $\sigma_1$ and $\sigma_2$ in our case).

Consider a portfolio consisting from this financial instrument,
$(-\delta_{1})$ units of the first asset and $(-\delta_{2})$ units
of the second asset. The price of the portfolio $\Pi(t)$ at a moment
$t$ is
$$
\Pi(t) = V - \delta_{1}S_{1} - \delta_{2}S_{2}.
$$

In this case, using the $\rm It\hat{o}$ formula, one can find the
law for the price of the financial instrument $V(S_1, S_2, t)$:
$$
dV = \left( V^{'}_{t} + \mu_{1}S_{1}V^{'}_{S_{1}} +
\mu_{2}S_{2}V^{'}_{S_{2}}
      + \frac{1}{2}\sigma^{2}_{1}S^{2}_{1}V^{''}_{S_{1}S_{1}}
      + \frac{1}{2}\sigma^{2}_{2}S^{2}_{2}V^{''}_{S_{2}S_{2}}
      + \sigma_{1}\sigma_{2}S_{1}S_{2}V^{''}_{S_{1}S_{2}} \right) dt +
$$
$$
     + \left( \sigma_{1}S_{1}V^{'}_{S_{1}} + \sigma_{2}S_{2}V^{'}_{S_{2}} \right) dW.
$$

A change of the portfolio price is written as
 $d\Pi = dV -
\delta_{1}dS_{1} - \delta_{2}dS_{2}$, the arbitrage-free condition
leads to the condition   $d\Pi = r(t)\Pi dt = r(t)\left( V -
\delta_{1}S_{1} - \delta_{2}S_{2}\right) dt$, where $r(t)$ - is the
spot interest rate. In what follows we set $r(t) \equiv r = const$.
We equate the right hand sides of  both expressions for  $d\Pi$ and
substitute the expression for $dV$. Thus, we get
$$
( V^{'}_{t} + \mu_{1}S_{1}V^{'}_{S_{1}} + \mu_{2}S_{2}V^{'}_{S_{2}}
      + \frac{1}{2}\sigma^{2}_{1}S^{2}_{1}V^{''}_{S_{1}S_{1}}
      + \frac{1}{2}\sigma^{2}_{2}S^{2}_{2}V^{''}_{S_{2}S_{2}}
      + \sigma_{1}\sigma_{2}S_{1}S_{2}V^{''}_{S_{1}S_{2}}
      - \delta_{1} \mu_{1} S_{1} - \delta_{2} \mu_{2} S_{2}) dt
$$
$$
      + \left(\sigma_{1}S_{1}V^{'}_{S_{1}} + \sigma_{2}S_{2}V^{'}_{S_{2}}
      - \delta_{1} \sigma_{1} S_{1} - \delta_{2} \sigma_{2} S_{2}\right) dW
      = r\left( V - \delta_{1}S_{1} - \delta_{2}S_{2}\right) dt.
$$

Equate coefficients at $dt$ and $dW$ with zero, we get two
equations:
$$
( V^{'}_{t} + \mu_{1}S_{1}V^{'}_{S_{1}} + \mu_{2}S_{2}V^{'}_{S_{2}}
      + \frac{1}{2}\sigma^{2}_{1}S^{2}_{1}V^{''}_{S_{1}S_{1}}
      + \frac{1}{2}\sigma^{2}_{2}S^{2}_{2}V^{''}_{S_{2}S_{2}}
      + \sigma_{1}\sigma_{2}S_{1}S_{2}V^{''}_{S_{1}S_{2}}
      -
$$
$$
     - \delta_{1} \mu_{1} S_{1} - \delta_{2} \mu_{2} S_{2})
     = r( V - \delta_{1}S_{1} - \delta_{2}S_{2}),
$$
$$
\sigma_{1}S_{1}V^{'}_{S_{1}} + \sigma_{2}S_{2}V^{'}_{S_{2}}
         - \delta_{1} \sigma_{1} S_{1} - \delta_{2} \sigma_{2} S_{2} = 0.
$$
The second equation gives
\begin{equation}
\label{delta1} \delta_{1} = V^{'}_{S_{1}} +
\frac{\sigma_{2}S_{2}}{\sigma_{1} S_{1}}V^{'}_{S_{2}}
         - \delta_{2} \frac{\sigma_{2}S_{2}}{\sigma_{1} S_{1}}
\end{equation}
Substitute this expression for $\delta_{1}$ in the first equation.
We get
$$
V^{'}_{t} + \mu_{1}S_{1}V^{'}_{S_{1}} + \mu_{2}S_{2}V^{'}_{S_{2}}
      + \frac{1}{2}\sigma^{2}_{1}S^{2}_{1}V^{''}_{S_{1}S_{1}}
      + \frac{1}{2}\sigma^{2}_{2}S^{2}_{2}V^{''}_{S_{2}S_{2}}
      + \sigma_{1}\sigma_{2}S_{1}S_{2}V^{''}_{S_{1}S_{2}} - \delta_{2} \mu_{2} S_{2}
      -
$$
$$
     - \mu_{1} S_{1}\left(V^{'}_{S_{1}} + \frac{\sigma_{2}S_{2}}{\sigma_{1} S_{1}}V^{'}_{S_{2}}
     - \delta_{2} \frac{\sigma_{2}S_{2}}{\sigma_{1} S_{1}}\right)
     = r\left( V - S_{1}\left(V^{'}_{S_{1}} + \frac{\sigma_{2}S_{2}}{\sigma_{1} S_{1}}V^{'}_{S_{2}}
     - \delta_{2} \frac{\sigma_{2}S_{2}}{\sigma_{1} S_{1}}\right) - \delta_{2}S_{2}\right).
$$
It can be readily shown that
\begin{multline}
\label{ur1}
  V^{'}_{t} + \frac{1}{2}\sigma^{2}_{1}S^{2}_{1} V^{''}_{S_{1}S_{1}}
  + \frac{1}{2}\sigma^{2}_{2}S^{2}_{2} V^{''}_{S_{2}S_{2}}
  + \sigma_{1}\sigma_{2}S_{1}S_{2} V^{''}_{S_{1}S_{2}}
  + r S_{1} V^{'}_{S_{1}}
\\
  + S_{2} \left(\mu_{2} - \mu_{1} \frac{\sigma_{2}}{\sigma_{1}} + r \frac{\sigma_{2}}{\sigma_{1}}\right) V^{'}_{S_{2}}
  + \delta_{2} S_{2} \left(\mu_{1}\frac{\sigma_{2}}{\sigma_{1}} - \mu_{2}  - r \frac{\sigma_{2}}{\sigma_{1}} + r\right)
  - r V = 0.
\end{multline}

Let us note that for the construction of the arbitrage hedging
strategy it is important that the coefficient at $\delta_{2}$ in
equation~\eqref{ur1} does not vanish. Thus, proving the
arbitrageness of the market we show by contradiction that for an
arbitrage-free market $\mu_{1}\frac{\sigma_{2}}{\sigma_{1}} -
\mu_{2}  - r \frac{\sigma_{2}}{\sigma_{1}} + r =0.$ This means that
at a arbitrage-free market the costs of risk
$\frac{\mu_i-r}{\sigma_i}$ for assets $S_1$ and $S_2$ coincide. This
unobvious fact can be proved differently by means of martingale
approach \cite{Bjork}.

We note that if one sets  $\delta_2=0$ (excluding a dependence of
$\Pi$ on $S_2$), than  \eqref{ur1} takes the form of the standard
Black-Scholes equation  (\cite{BS}).

Assume that a buyer of the option does not know that the seller is
going to get an additional asset  to take part in hedging and
therefore he is oriented to the option price, found by the  standard
Black-Scholes formula. Therefore we set an initial-boundary problem
for equation \eqref{ur1}, imitating  the Cauchy problem for the
standard Black-Scholes equation. We denote the solution of the
latter one as  $\bar V(t,S_1)$ and set the "`final"' equation $\bar
V(S_{1}, T) = \left( S_{1} - X\right)^{+}$, where $\left( S_{1} -
X\right)^{+} = \max \left(S_1 - X, \; 0\right), \quad X = const >
0$. As follows from explicit formula for the solution of this
problem, at any moment of time  $t\in [0,T]$ we have $\bar V(0,
t)=0,\, $ and $\bar V(S_{1}, t)=S_1-X
e^{-r(T-t)}(1-o(\frac{1}{S_1})),\,$ at $S_1\to +\infty$. Let us
choose a large positive number $K_+$ and large by modulus negative
number $K_-$. We denote
\begin{equation}\label{S}S_\pm=e^{\frac{ K_\pm}{\alpha}-c(T-t)},\end{equation} where $\alpha>0$ and $c$
are constants, which will be chosen later, $t\in [0,T].$ It is clear
that  $S_-\to 0$ and $S_+\to +\infty$ as $|K_\pm|\to \infty$. We
choose functions $g_\pm(S_\pm,t)=\bar V(S_\pm, t).$ Thus, $g_-(S_-,
t)=o(S_-),\, $ as $\,S_-\to 0$ and $\bar V(S_+, t)=S_+-X
e^{-r(T-t)}(1-o(\frac{1}{S_+})),\,$ as $S_+\to +\infty$.

So, for any how many large by modulus positive number $K_+$,
negative number $K_-$ and any $S_2>0$ we get the initial-boundary
problem for equation \eqref{ur1}:
\begin{equation}
\label{nu1} V(S_{1}, S_{2}, T) = \left( S_{1} - X\right)^{+},
\end{equation}
\begin{equation}
\label{ku1} V(S_-, S_2, t) = g_-(S_-,t), \quad V(S_+, S_2, t) =
g_+(S_+,t).
\end{equation}
 We can justify the change from the semi-axis $S_1>0$ to the segment $[S_-,S_+]$ by including in the terms of
 contract a condition on cancelation of the contract  if the price oversteps the limits specified beforehand
 within a time  $t\in
[0,T]$ (the price corridor can be arbitrary large).

\section{Reducing  \eqref{ur1} to semi-linear parabolic equation}

Let us perform several changes of independent and dependent
variables of problem \eqref{ur1}-\eqref{ku1} and reduce it to an
initially-boundary problem for the heat equation. We do not write
the respective initial-boundary problem at every step, only list the
changes performed. We note that these change analogous in outline to
the changes that reduce the Black-Scholes equation to the heat
equation. Nevertheless, there are some distinctions.
\\
%{\bf 1.}

We make the change of the time direction $ \tau = T - t$; the change
of independent variables  $x_1 = \alpha_1 \ln S_{1}$, $\quad x_2 =
\alpha_2 \ln \frac{S^{\sigma_{1}}_{2}}{S^{\sigma_{2}}_{1}}$, where
$\alpha_1, \alpha_2 > 0$ are arbitrary constants; the change of
dependent variable  $V(x_1, x_2, \tau) = e^{-r\tau} U(x_1, x_2,
\tau)$; the shift $y_{1} = x_{1} +  c_{1} \tau, y_{2} = x_{2} +
c_{2} \tau,$ where $c_{1} = \alpha_{1} \left(r -
\frac{1}{2}\sigma^{2}_{1} \right)$ ш
    $c_{2} = \alpha_{2} \left(\mu_{2} \sigma_{1} - \mu_{1} \sigma_{2} +
    \frac{1}{2}\sigma_{1}\sigma_{2}(\sigma_{1} - \sigma_{2})\right).$

 If in \eqref{S} we choose  $\alpha=\alpha_1$ and $c=c_1/\alpha,$ then
 we obtain the following initial-boundary problem
\begin{equation}
\label{ursved}
\begin{array}{ll}
  \frac{1}{2}\sigma^{2}_{1} \alpha^{2}_{1} U^{''}_{y_{1}y_{1}}
  + \delta_{2}e^{r\tau + \frac{y_{2} - c_{2}\tau}{\alpha_{2}\sigma_{1}} + \frac{(y_{1} - c_{1}\tau)\sigma_{2}}{\alpha_{1}\sigma_{1}}}
    \left(\mu_{1}\frac{\sigma_{2}}{\sigma_{1}} - \mu_{2} - r \frac{\sigma_{2}}{\sigma_{1}} + r\right)
  = U^{'}_{\tau}, \\
U(y_{1}, y_{2}, 0) = \left( e^{\frac{\left.y_{1}\right|_{\tau = 0}}{\alpha_{1}}} - X\right)^{+}, \quad y_1 \in [K_-, K_+], \; y_2 \in \mathbb R, \\
U(K_-, y_{2}, \tau) =  e^{r\tau} g_-(S_-, \tau), \qquad U(K_+,
y_{2}, \tau) = e^{r\tau} g_+(S_+, \tau).
\end{array} %\right.
\end{equation}

We note that we can consider the variable $y_{2}$ as a parameter,
since in equation \eqref{ursved} there is not derivatives with
respect to it, but the dependence of  $y_{2}$ remains.

\indent We introduce the notations $\varepsilon^{2} =
\frac{1}{2}\sigma^{2}_{1}\alpha^{2}_{1}$, $\quad U_{0}(y_{1},
\varepsilon) =$ $ \left( e^{\frac{\left.y_{1}\right|_{\tau =
0}}{\alpha_{1}}} - X\right)^{+}$,\\ $\quad F(y_{1}, \tau,
\varepsilon) = - e^{r\tau + \frac{y_{2} -
c_{2}\tau}{\alpha_{2}\sigma_{1}} + \frac{(y_{1} -
c_{1}\tau)\sigma_{2}}{\alpha_{1}\sigma_{1}}}
                   \left(\mu_{1}\frac{\sigma_{2}}{\sigma_{1}} - \mu_{2} - r \frac{\sigma_{2}}{\sigma_{1}} +
                   r\right).$
                   Since the variable  $\delta_{2}$, corresponding to a share of the second asset
                   in the riskless  portfolio, can be chosen
                   arbitrary, then we set $ \delta_{2} = U(U - A)(U - B) \left(F(y_{1}, \tau,
\varepsilon)\right)^{-1}$, where $A$ and $B$ are certain constants
to be defined below. Such choice of  $\delta_{2}$ leads to a problem

\begin{equation}
\label{ursved3}
\begin{array}{ll}
\varepsilon^{2}U^{''}_{y_{1}y_{1}} - U^{'}_{\tau} = f(U),  \\
U(y_{1}, 0, \varepsilon)  = U_{0}(y_{1}, \varepsilon), \quad y_1 \in
\mathbb R,\\ U(K_-, y_{2}, \tau) =  e^{r\tau} g_-(S_-, \tau), \quad
U(K_+, y_{2}, \tau) = e^{r\tau} g_+(S_+, \tau).
\end{array}
\end{equation}
where $f(U) = U(U - A)(U - B)$.

Let us approximate the boundary conditions taking into account the
fact that the value of  $K$ is large and  $\tau$ is bounded. Note
that $e^{r\tau} g_-(S_-,\tau)\to 0$ as $|K_-|\to\infty$ and $
e^{r\tau} g_+(S_+, \tau)=e^{\frac{K}{\alpha}+\frac{\sigma^2
\tau}{2}}-X.$ Then under additional assumption  $\sigma^2 T\ll 1$ we
exclude the dependence on time in the boundary conditions:
\begin{equation}
\label{bc}
\begin{array}{ll}
 U(K_-, y_{2}, \tau) = 0,
\quad U(K_+, y_{2}, \tau) = e^{\frac{K_+}{\alpha_1}} - X.
\end{array}
\end{equation}
Since in the expression $\varepsilon^{2} =
\frac{1}{2}\sigma^{2}_{1}\alpha^{2}_{1}$ the parameter $\alpha_1$ is
arbitrary, we can make $\varepsilon$ as small as we want.

\section{Conditions for formation of the step-like contrast structure}

Let us  outline known results on conditions of formation of the
step-like contrast structure \cite{BKN}. Consider the following
initial-boundary problem
\begin{equation}
\label{krzad}
%\left\{
\begin{array}{ll}
\varepsilon^{2} u^{''}_{xx} - u^{'}_{t} = f(u,\; x,\; \varepsilon), \quad (x, t) \in D\times(0,\; +\infty),  \\
u(a,\; t,\; \varepsilon) = g_{a}, \quad u(b,\; t,\; \varepsilon) = g_{b}, \quad t \in (0,\; + \infty),  \\
u(x, 0, \varepsilon) = u_{0}(x,\; \varepsilon), x \in \bar{D},
\end{array}
%\right.
%\eqno(9)
\end{equation}
where $\varepsilon > 0$ is a small parameter, $D \equiv (a, b)$,
$g_{a}$ ш $g_{b}$ are constants.
\\
Assume that the function  $f$ satisfies the following conditions.
\\
$(\bf A1).$ There exist functions $\bar{\omega}$ and $\hat{\omega}$
from $C^{2}(\bar{D})$ such that  $\bar{\omega} < \hat{\omega}, x \in
\hat{D},$ and in the domain  $\Omega = \{(u, x): \bar{\omega}
\leqslant u \leqslant \hat{\omega}, \; x \in \hat{D} \}$ the
function $f(u,\; x,\; 0)$ vanishes only on the curves  $u =
\varphi_{i}(x), i = 0, 1, 2,$ moreover,
$$
\bar{\omega} < \varphi_{1}(x) < \varphi_{0}(x) \equiv 0 <
\varphi_{2}(x) < \hat{\omega}, \quad x \in \hat{D},
$$
$$
f_{u}(\varphi_{i}(x),\; x,\; 0) > 0,  i = 1, 2; \quad
f_{u}(\varphi_{0}(x),\; x,\; 0) < 0, \quad x \in \hat{D};
$$
Assume that  $f(u,\; x,\; \varepsilon)$ is sufficiently smooth
function in the domain $\Omega_{1} \times [0, \varepsilon_{0}]$,
where $\Omega_{1}$ contains $\Omega$ and $\varepsilon_{0}
> 0$ is an arbitrary number.
\\
We  set  $\varphi_0 \equiv 0$ for the sake of simplicity only. It is
not difficult to reformulate all results for the case $\varphi_0
\not \equiv 0$.
\\
We introduce a function  $J(x) =
\int\limits^{\varphi_{2}(x)}_{\varphi_{1}(x)} f(u,\; x, \; 0) du$
and make the following assumption.
\\
$(\bf A2).$ There exists a point  $x_{0} \in D$ such that $J(x_{0})
= 0, \; \frac{dJ}{dx}(x_{0}) < 0$.
\\
$(\bf A3).$ The following inequalities take place: $\varphi_{1}(a) <
g_{a} < \varphi_{2}(a), \quad \varphi_{1}(b) < g_{b} <
\varphi_{2}(b)$,
$$
\int\limits^{y}_{\varphi_{1}(a)} f(u,\; a,\; 0) du > 0, \; y \in
(\varphi_{1}(a), \; g_{a}], \quad \int\limits^{y}_{\varphi_{2}(b)}
f(u,\; b,\; 0) du > 0, \; y \in (g_{b}, \; \varphi_{2}(b)].
$$

Under conditions $(\bf A1) - (\bf A3)$ for sufficiently small
$\varepsilon$ there exists a stationary solution $u_{s}(x,
\varepsilon)$ to the boundary problem having an internal  transition
layer in a neighborhood of a point  $x_{0}$ such that
\begin{equation}
\label{pred} \lim_{\varepsilon \to 0} u_{s}(x, \varepsilon) =
\left\{
\begin{array}{ll}
\varphi_{1}(x), \quad x \in (a, x_{0})  \\
\varphi_{2}(x), \quad x \in (x_{0}, b).
\end{array} \right.
\end{equation}
The solution of this kind are called contrast step-like structure
 (CSLS).
\\

It is known that under conditions  $(\bf A1) - (\bf A3)$ the CSLT
solution  $u_{s}(x, \varepsilon)$ is an asymptotically stable
solution to the boundary problem. There arises a question on a set
of initial values  $u_{0}(x, \varepsilon)$, which lead to a
formation of the contrast structure $u_{s}(x, \varepsilon)$ as $t
\rightarrow +\infty$. In other words, which is a domain of influence
of this solution? Let us give the definition of the global domain of
influence according to  \cite{BKN}.
\\

Let the boundary problem have a stationary solution
$u_{\varepsilon}(x) \in C^{2}(\bar{D})$ at $\varepsilon \in (0,
\varepsilon ']$, where $\varepsilon ' > 0$ is a certain number.
\\
\begin{definition} We call  $G(u_{\varepsilon})$ the global domain of influence
of a stationary solution  $u_{\varepsilon}(x)$ of the boundary
problem if  $G(u_{\varepsilon})$ contains functions  $u_{0}(x,
\varepsilon)$ having the following property: there exists
$\varepsilon '' \in (0, \varepsilon ']$ such that at $\varepsilon
\in (0, \varepsilon '']$ there is a solution $u_{\varepsilon}(x, t)
\in C^{1,0}(\bar{D} \times [0, +\infty)) \cap C^{2,1}(\bar{D} \times
(0, +\infty))$ of the initial-boundary problem and $\lim_{t \to
+\infty} ||u_{\varepsilon}(x, t) - u_{\varepsilon}(x)||_{C(\bar{D})}
= 0$.
\end{definition}

Let  $f$ satisfy additional conditions:
\\
$(\bf A4).$ The set of all points  $x$ such that $J(x) = 0$ consists
of finite number of segments or points.
\\
$(\bf A5)$. $u_{0}(x, \varepsilon) \equiv u_{0}(x) \in
C^{2}_{B}(\bar{D}) \equiv \{v(x) \in C^{2}(\bar{D}): v(a) = g_{a},
\; v(b) = g_{b} \}$ ш
$$
\varphi_{1}(x) \leqslant u_{0}(x) \leqslant \varphi_{2}(x), x \in
\bar{D}.
$$
$(\bf A6)$. $\exists x^{(-)} \in (a, x_{0})$ and $\exists x^{(+)}
\in (x_{0}, b)$ such that
$$
u_{0}(x^{(-)}) < \varphi_{0}(x) \mbox{ ш } u_{0}(x) < \varphi_{0}(x)
\mbox{ тю тёхї Єюўърї } x \in[a, x_{0}), \mbox{ уфх } J(x) \leqslant
0,
$$
$$
u_{0}(x^{(+)}) > \varphi_{0}(x) \mbox{ ш } u_{0}(x) > \varphi_{0}(x)
\mbox{ at all points  } x \in(x_{0}, b], \mbox{ where } J(x)
\geqslant 0.
$$
\indent The main result of \cite{BKN}  is the following theorem.
\begin{theorem}
Let conditions $(\bf A1) - (\bf A3)$ hold. Then for sufficiently
small  $\varepsilon$ there exists a stationary solution $u_{s}(x,
\varepsilon)$ of the boundary problem \eqref{krzad} from
$C^{2}(\bar{D})$, having form of the contrast step-like structure,
satisfying limit equation  \eqref{pred}.

Moreover, let  condition $(\bf A4)$ hold. Then the following is
true:
\\
1. If a function $u_{0}(x, \varepsilon) \equiv u_{0}(x)$ satisfies
$(\bf A5), (\bf A6)$, then it  falls in $G(u_{s})$.
\\
2. If $\exists \varepsilon_{1} > 0$ such that at $\varepsilon \in
(0, \varepsilon_{1}]$ a function  $u_{0}(x, \varepsilon)$ satisfies
condition  $(\bf A5)$ and there exist functions $\bar{u}_{0}(x)$ and
$\hat{u}_{0}(x)$, satisfying $(\bf A5), (\bf A6)$ and such that
$$
\bar{u}_{0}(x) < u_{0}(x, \varepsilon) < \hat{u}_{0}(x), \quad x \in
\bar{D}, \quad \varepsilon \in (0, \varepsilon_{1}],
$$
then $u_{0}(x, \varepsilon)$ belongs to $G(u_{s})$.
\end{theorem}

\section{Formation of SLCS in problem~\eqref{ursved3}} \indent

We re-formulate  problem \eqref{ursved3} with boundary conditions,
changed to \eqref{bc}  (we write  $x$ instead of $y_1$ and $u$
instead of $U$):
\begin{equation}
\label{urksts}
\begin{array}{ll}
\varepsilon^{2} u^{''}_{xx} - u^{'}_{t} = f(u),  \\
u(x, 0) = \left( e^{\frac{x}{\alpha_{1}}} - X\right)^{+}, \\
u(a, t) = 0,  \\
u(b, t) = e^{\frac{K_+}{\alpha_1}} - X,
\end{array}
\end{equation}
where $f(u) = u(u - A)(u - B), \quad 0 < A < B,\quad t \in [0, T],
\quad x \in [a, b],\,a=K_-,\,b=K_+$.
\\

Let us apply the results of the CSLS theory  outlined in Sec. 4 to
problem \eqref{urksts}. Condition  $(\bf A1)$ holds evidently. From
condition $(\bf A2)$ we get a relation between constants  $A$ and
$B$. Since $\int\limits^{B}_{0} f(u) du = \int\limits^{B}_{0} u(u -
A)(u - B) du = 0$ , then  $B = 2A$.

Thus, the condition $(\bf A4)$ holds since  $J(x) \equiv 0$ for all
segment  $[a, b]$. It remains to find $A$ to satisfy $(\bf A3)$. We
take  $A = \frac{u(K_+,\; t)}{2} = \frac{e^{\frac{K_+}{\alpha_1}} -
X}{2}$.
\\
The point of transfer can be found as
$$
x^0 = a+(b - a)\frac{\sqrt{f_u(\varphi_2)}}{\sqrt{f_u(\varphi_2)} +
\sqrt{f_u(\varphi_1)}}
    = K_-+ (K_+-K_-)\frac{\sqrt{f_u(2A)}}{\sqrt{f_u(2A)} + \sqrt{f_u(0)}}
 = \frac{K_-+ K_+}{2}
$$
(see Butuzov, Vassilieva).

Thus, the contrast structure in the stationary problem has a form
$$
\lim_{\varepsilon \to 0} u(x, \varepsilon) = \left\{
\begin{array}{ll}
0, \mbox{ яЁш } K_- < x < 0,  \\ \\
e^{\frac{K_+}{\alpha_1}} - X, \mbox{ яЁш } 0< x < K_+.
\end{array} \right.
$$

Condition  $(\bf A5)$ holds due to the relation $u(x, 0) \equiv
\left( \varphi_2(x), 0 \right)^{+}$, and condition $(\bf A6)$ holds
due to  $u(x, 0) \equiv \frac12 \varphi_0(x)$. Since all conditions
of Theorem 1 take place, then in the problem \eqref{urksts} a
contrast step-like structure arises.

\section{Arbitrage hedging strategy}

 We note that we can choose the values  $K_-,\; K_+$ arbitrary. Therefore, increasing $S_+$
 one can always shift the transition point   $S_0 = \sqrt{S_+ S_-}$ a little to the right of
  the strike price $X$. This means that we can chose such hedging strategy
   $(\delta_1, \delta_2)$, that the option price can be negligibly small initially.

The option price, found by the classical Black-Scholes formula, it
is greater initially than at the moment $T$. But if we apply the
hedging strategy involving the second asset, then we get that the
initial option price  (at $S < S_0$) is negligible comparing with
its price at the moment of exercise. This gives evidently a
possibility  of arbitrage upon conclusion of contract. Thus, we get
a contradiction with the assumption with a non-arbitrageness of the
market.

The hedging strategy  $(\delta_1, \delta_2)$ has a form:
$$
\delta_{1} = V^{'}_{S_{1}} + \frac{\sigma_{2}S_{2}}{\sigma_{1}
S_{1}}V^{'}_{S_{2}}
         - \delta_{2} \frac{\sigma_{2}S_{2}}{\sigma_{1} S_{1}},\qquad
\delta_2 = V(V - (S_+ - X)/2)(V - (S_+ - X)) \frac{1}{S_2 (\mu_2 -
\mu_1 \frac{\sigma_2}{\sigma_1})}.
$$

\section{Numerical solution of problem~\eqref{urksts}}

Let us construct the contrast structure in our problem numerically.
In particular, we understand how quick the structure forms. We use
the Crank - Nicolson method and the marching. Recall that we seek
for the solution to the problem \eqref{urksts} in the domain
$[K_-,K_+] \times [0,T]$.
\\

We choose the following parameters:
 $S_- = 0.1, \;
S_+ = 100,$  $N = 100, \tau = 2 \cdot 10^{-4}, \alpha_1 = 1,
\sigma_1 = 0.02, X = 20$. Then the transfer point is  $S_0 = 31.6$,
the size of the step is $A = 40$.
\\

Pic.1 presents the "final " function  $V(S_1, S_2, T)$ (solid line),
and the numerical solution $V(S_1, S_2, t)$ at $t = 0$ (dashed
line). Even at  $T = 0.25$ (the exercise time equal to 3 months) the
graph of solution is step-like.
\\
Pic.2 presents the graph of the same function  $V = V(S_1, t)$ яЁш
$t = 0$ (points) and at  $t = T$ (solid line) by hedging without
using the second asset  (х.g. $\delta_2 = 0$) as in classical
Black-Scholes  model.

\begin{figure}[h]
\begin{minipage}{0.5\columnwidth}
\centerline{\includegraphics[width=1\columnwidth]{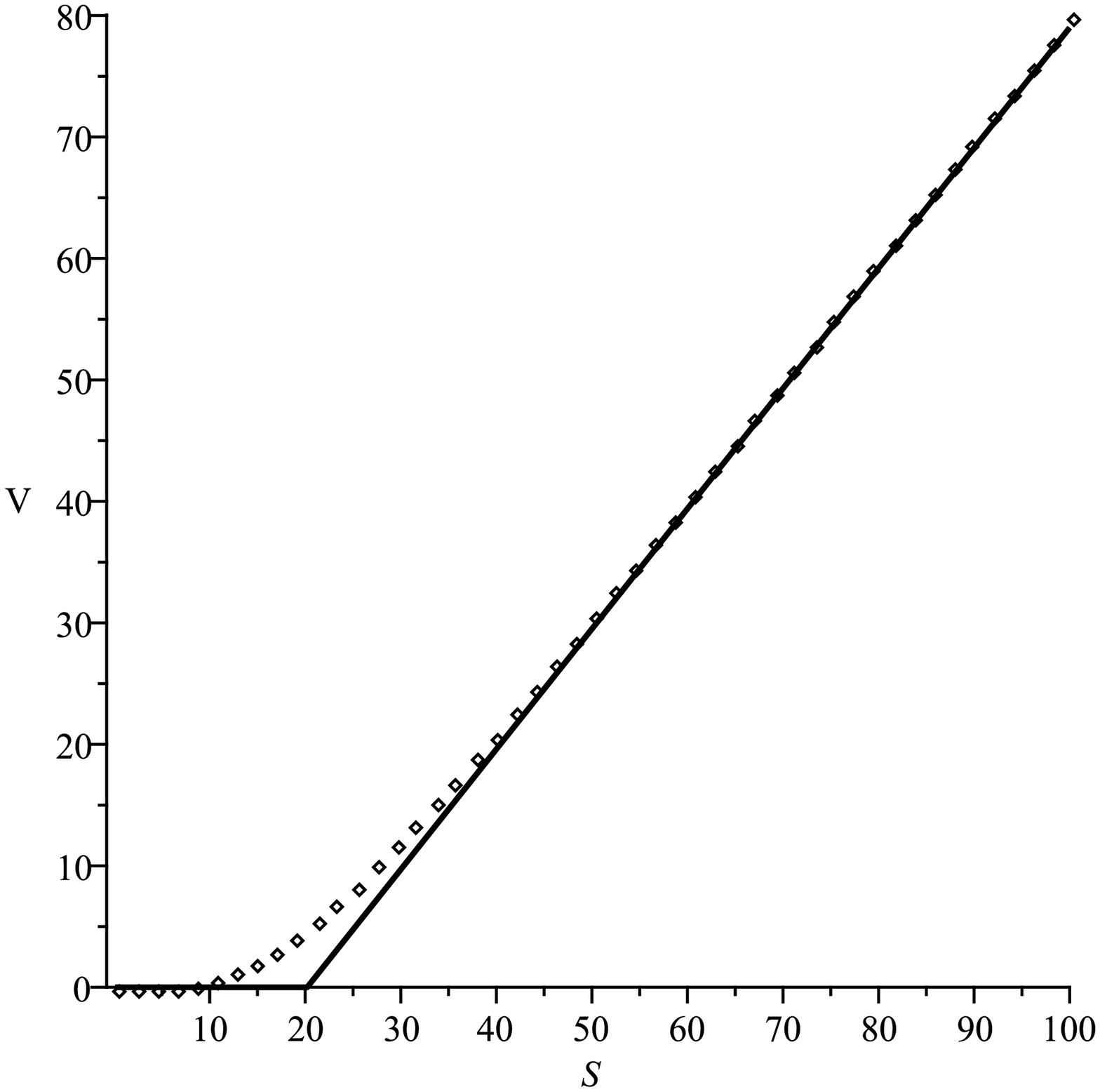}}
%\caption{Pic.1}
\end{minipage}%
%\end{figure}%
%\begin{figure}[h]
\begin{minipage}{0.5\columnwidth}
\centerline{\includegraphics[width=1\columnwidth]{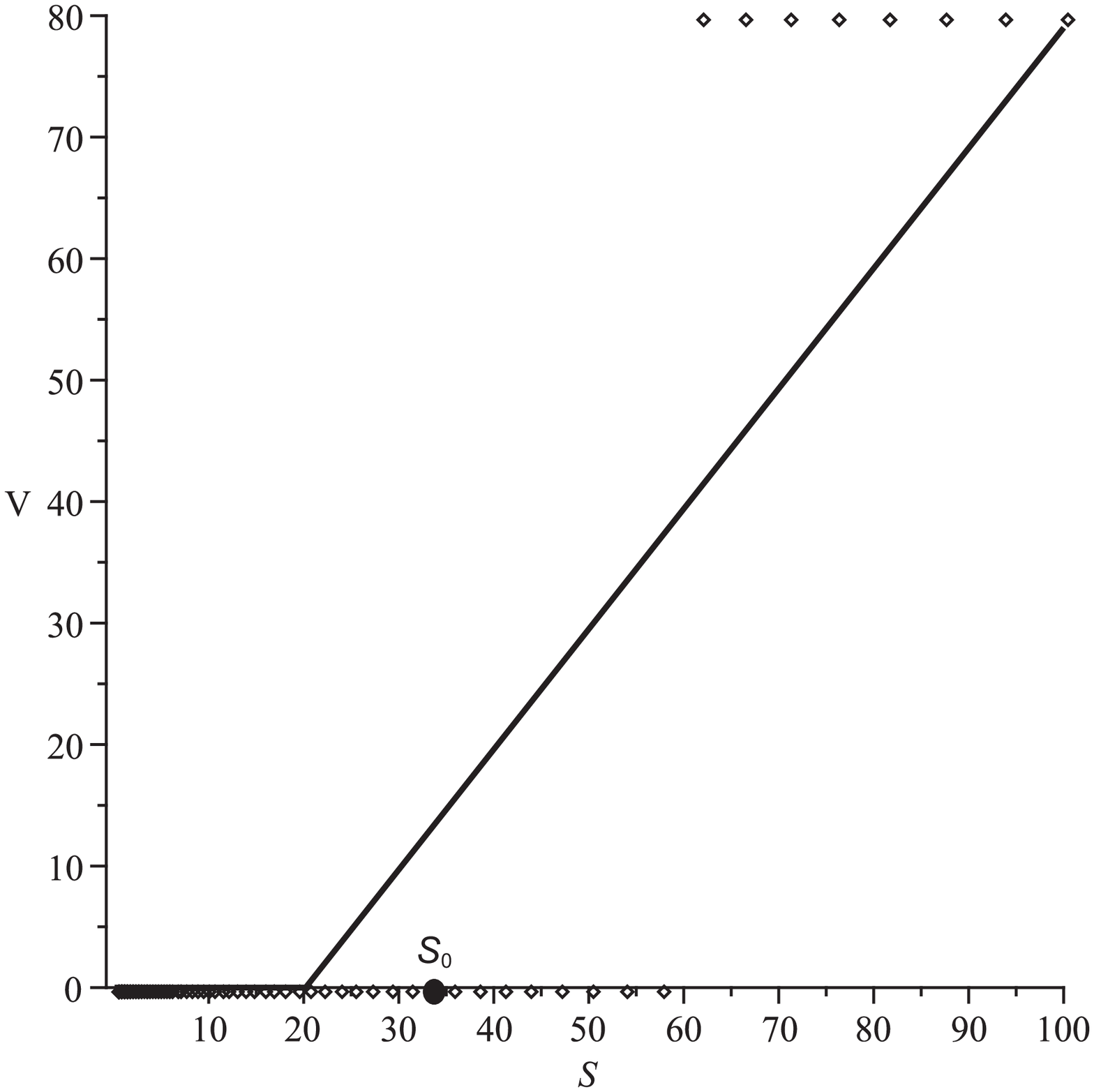}}
%\caption{Pic.2}
\end{minipage}
\end{figure}

%\def\kscale{0.8}
%\makepictwo{Graphic2.eps}{╨шё.\ 1.}{gr1.eps}{╨шё.\ 2. }

\section{"Volatility smile" as a manifestation of arbitrageness of market}

Below we present the graphs of the option price in dependence on the
strike price $X$ (fixed spot price and the time of exercise) for the
arbitrage-free Black-Scholes hedging strategy and arbitrage hedging
strategy considered in this paper (Pic.3). Pic.4 presents the graphs
for the dependence of volatilities on the strike price computes
basing on the Black-Scholes formula for different times of exercise.
The graphs clearly demonstrate the "skew smile", moreover, the skew
increases as the exercise time gets smaller, as is in compliance
with observational data (e.g. \cite{Derman}). It is interesting that
for explanation of this phenomenon we do not need to engage the idea
of stochastic volatility as they usually do. We note that the
phenomena of "smile" of volatility near the strike price is observed
since 1987, where the amendment to the Glass-Steagall Act allowed to
invest the bank capitals in derivatives. Since the asset and its
derivative are correlated, it inevitably gives an arbitrage
possibility and therefore the main conditions for the Black-Scholes
formula can not be satisfied.
\begin{figure}[h]
%\begin{minipage}{1\columnwidth}
\centerline{\includegraphics[width=1\columnwidth]{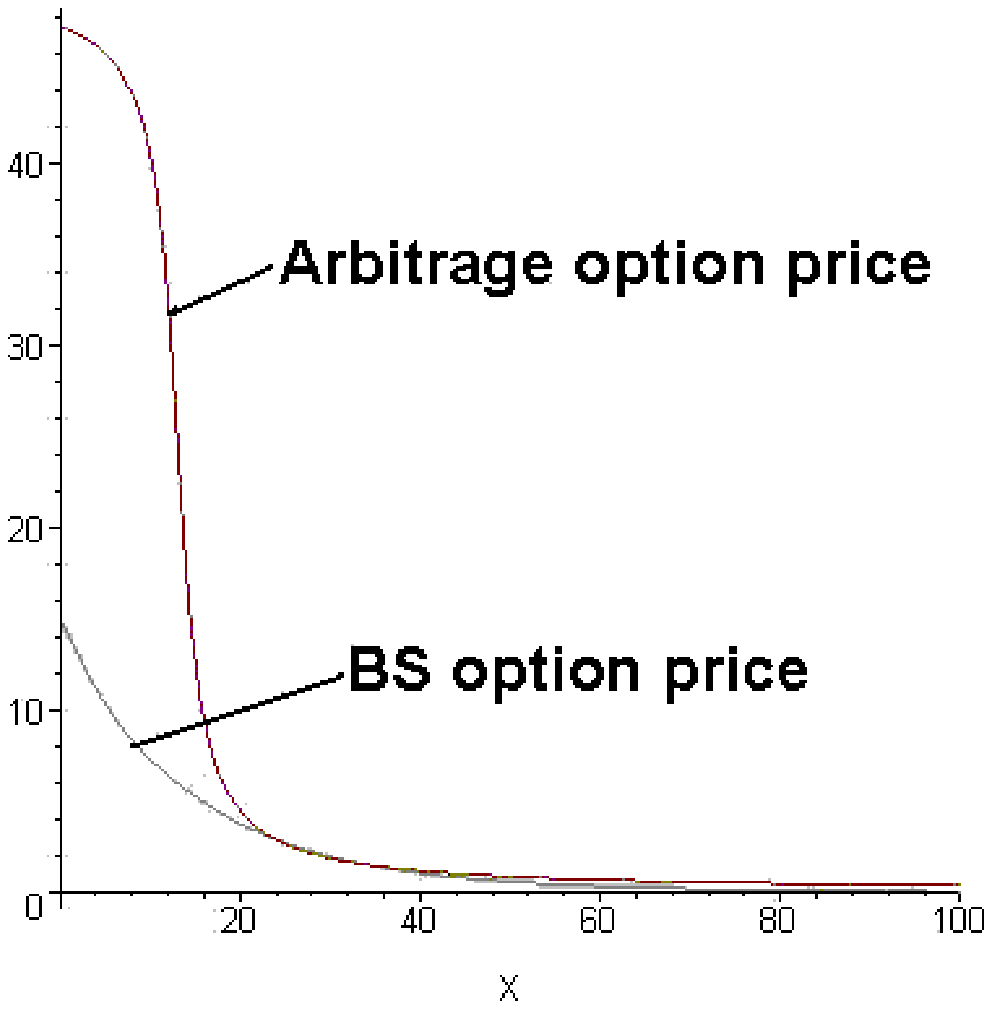}}
%\caption{Pic.1}
%\end{minipage}%
\end{figure}%
\begin{figure}[h]
%\begin{minipage}{1\columnwidth}
\centerline{\includegraphics[width=1\columnwidth]{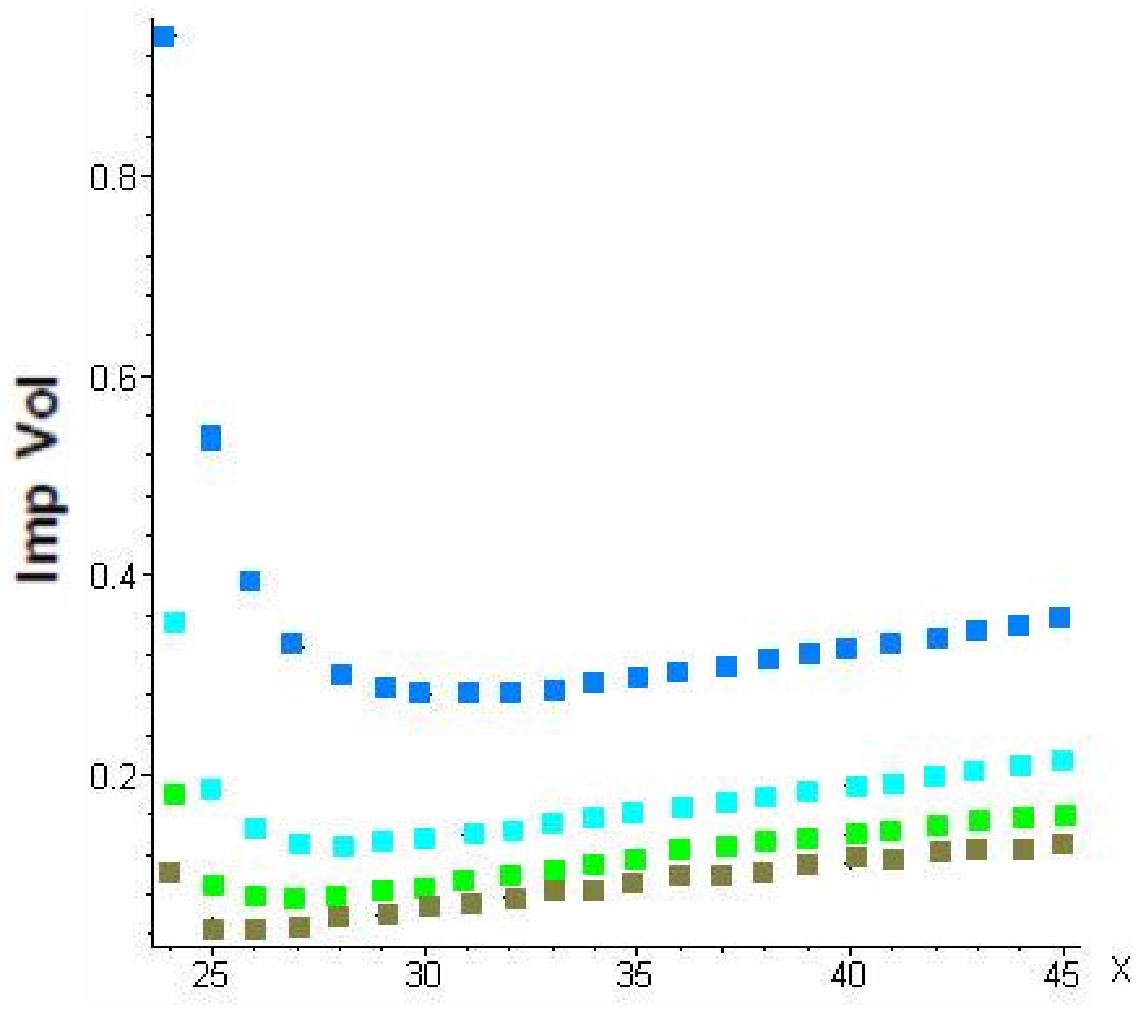}}
%\caption{Pic.2}
%\end{minipage}
%\caption{Density and velocity, $u_2<0.$} \label{Fig2}
\end{figure}

\end{document}